\documentclass[twoside]{dis07}
\usepackage[latin1]{inputenc}
\usepackage[dvips]{graphicx,epsfig,color}
\usepackage{wrapfig,rotating}
\usepackage{amssymb,amsmath,array}

\begin{document}
\title{Recent developments in small-$x$ physics}

%***********************************************************************
% AUTHORS INFORMATION AREA
%***********************************************************************
\author{Arif I. Shoshi$^1$
%
% Optional short acknowledgment: remove next line if non-needed
\thanks{The author acknowledges financial support by the DFG under contract 92/2-1.}
%
% DO NOT MODIFY THE FOLLOWING '\vspace' ARGUMENT
\vspace{.3cm}\\
%
% Addresses and institutions (remove "1- " in case of a single institution)
1-Fakult{\"a}t f{\"u}r Physik, Universit{\"a}t Bielefeld, D-33501 Bielefeld,
Germany}

%School of First Author - Dept of First Author \\
%Address of First Author's school - Country of First Author's
%school
%
% Remove the next three lines in case of a single institution
%\vspace{.1cm}\\
%2- School of Second Author - Dept of Second Author \\
%Address of Second Author's school - Country of Second Author's school\\
%}
%***********************************************************************
% END OF AUTHORS INFORMATION AREA
%***********************************************************************

\maketitle

\begin{abstract}
  Recent theoretical progress in understanding high-energy scattering beyond the mean field approximation is reviewed. The role of Lorentz invariance and
  pomeron loops in the evolution, the relation between high-energy QCD and statistical physics
  and results for the saturation momentum and the scattering amplitude are
  discussed.
%the of 
%It is shown how to get resultsfor the
%  saturation momentum and the scattering amplitude beyond the mean field
%  approximation. 
%The importance of boost invariance of the small-$x$
%  evolution, the relation between high-energy QCD and statistical physics, the
%  role of pomeron loops and results for saturation momentum and the scattering
%  amplitude beyond the mean field approximation are discussed.
%emphasized.
%, the energy dependence of the saturation momentum and the
%  scaling behaviour of the scattering amplitude at high energy are discussed.
\end{abstract}

\section{Introduction}

The high-energy scattering of a dipole off a nucleus/hadron in the {\em mean
  field approximation} is decribed by the
BK-equation~\cite{Balitsky:1995ub+X}. The main results following from the
BK-equation are the geometric scaling behaviour of the scattering amplitude
and the roughly powerlike energy dependence of the saturation scale which are
both nicely supported by the HERA data.

The recent progress consists in understanding small-$x$ dynamics (near the
unitarity limit) {\em beyond the mean field approximation}, i.e., beyond the
BK-equation. A first step beyond the mean field approximation was done
in~\cite{Mueller:2004se} where the BFKL evolution in the scattering process
was enforced to satisfy natural requirements as unitarity limits and Lorentz
invariance. The result was a correction to the saturation scale and the
breaking of the geometric scaling at high energies.  Afterwards a relation
between high-energy QCD and statistical physics was found~\cite{Iancu:2004es}
which has clarified the physical picture of, and the way to deal with, the
dynamics beyond the BK-equation. It has been understood that {\em gluon number
  fluctuations} from one scattering event to another and the {\em
  discreteness} of gluon numbers, both ignored in the BK evolution and also in
the Balitsky-JIMWLK equations~\cite{Iancu:2003xm}, lead to the breaking of the
geometric scaling and to the correction to the saturation scale,
respectivelly.  New evolution
equations~\cite{Mueller:2005ut,Iancu:2004iy,Kovner:2005nq}, which describe
Pomeron loops, have been proposed to account for the above effects. Very
recently possible effects of Pomeron loops on various
observables~\cite{Kozlov:2006qw,Hatta:2006hs,Iancu:2006uc} have been studied
in case they become important in the range of collider energies.

%The recent progress consists in understanding small-$x$ dynamics (near the
%unitarity limit) {\em beyond the mean field approximation}, i.e., beyond the
%BK-equation. The first results beyond the mean field approximation, a
%correction to the energy dependence of the saturation scale and the breaking
%of the geometric scaling at high energies, have been obtained
%in~\cite{Mueller:2004se} by requiring for the QCD evolution in the scattering
%process to satisfy unitarity limits and Lorentz invariance. Soon after a
%relation between high-energy QCD and statistical physics was
%found~\cite{Iancu:2004es} which has provided a physical picture of, and a
%better way to deal with, the dynamics missed in the BK-equation. This work has
%clarified that the elements not taken into account in the BK-equation, namely,
%the {\em gluon number fluctuation} from one scattering event to another and
%the {\em discreteness} of gluon numbers, lead, respectivelly, to the breaking
%of the geometric scaling and to the correction to the saturation scale. Later
%on, new evolution equations~\cite{Mueller:2005ut,Iancu:2005nj,Kovner:2005nq}
%have been proposed aiming at the accomodation of gluon number fluctuations, or
%Pomeron loops, that are not included in the BK-equation, or the more
%sophisticated Balitsky-JIMWLK equations~\cite{Iancu:2003xm}. Very recently
%possible effects of Pomeron loops on
%observables~\cite{Kozlov:2006qw,Iancu:2006uc} have been studied, in case they
%become important in the range of collider energies.

In the following sections I will show the recent developments in some detail
by considering equations and results in and beyond the mean
field approximation.

\section{Mean field approximation}
Consider the high-energy scattering of a dipole of transverse size $r$ off a
target (hadron, nucleus) at rapidity $Y = \ln(1/x)$. The rapidity dependence
of the $T$-matrix in the mean field approximation is given by the BK-equation which has the schematic structure (transverse dimensions are suppressed) 
\begin{equation}
\partial_Y T = \alpha_s \left[ T -  T \ T \right] \ .
\end{equation}
The linear part of the BK equation, i.e., the BFKL equation, gives the growth
of $T$ with rapidity $Y$ whereas the non-linear term $T T$ tames the growth of
$T$ in such a way that the unitarity limit $T \leq 1$ is satisfied.

One of the main results following from the BK-equation is the {\em geometric
scaling} behaviour of the $T$-matrix~\cite{Stasto:2000er}  
\begin{equation}
T(r,Y) = T(r^2\,Q^2_s(Y)) \ ,
\label{eq:gs}
\end{equation}
where $Q_s(Y)$ is the so-called {\em saturation momentum} defined such that
$T(r \simeq 1/Q_s,Y)$ be of ${\cal{O}}(1)$.  Eq.~(\ref{eq:gs}) means that
the $T$-matrix scales with a single quantity $r^2\,Q^2_s(Y)$ rather than
depending on $r$ and $Y$ separatelly. This behaviour implies a similar
scaling for the DIS cross section, $\sigma^{\gamma^{*}p}(Y,Q^2) =
\sigma^{\gamma^{*}p}(Q^2/Q^2_s(Y))$, which is supported by the HERA data.

Another important result that can be extracted from the BK-equation is the
rapidity dependence of the saturation momentum (leading-$Y$
contribution)\cite{Mueller:2002zm+X},
\begin{equation} 
Q^2_s(Y) = Q_0^2 \ \mbox{Exp}\left[\frac{2\alpha_s N_c}{\pi}
\frac{\chi(\lambda_0)}{1-\lambda_0} Y \right] \ ,  
\label{eq:Qrmf}
\end{equation}
where $\chi(\lambda)$ is the BFKL kernel and $\lambda_0 = 0.372$.  

%The
%exponent in (\ref{eq:Qrmf}) is known at NLO accuraccy and has the value
%$\lambda \simeq 0.3$ which is in good agreement with values obtained from fits
%to the HERA data.

The shape of the $T$-matrix resulting from the BK-equation is preserved in the
transition region from weak to strong scattering, $0 < T < 1$, with
rising $Y$ (front of the travelling wave): The saturation region at $r \gg 1/Q_s(Y)$
where $T \simeq 1$ however widens up, including smaller and smaller dipoles,
due to the growth of the saturation momentum. The situation changes, as we
will, once gluon number fluctuations are taken into account.

\section{Beyond the mean field approximation}
\label{sec:bmf}

\subsubsection{Lorentz invariance and unitarity requirements}

Let's start with an elementary dipole of size $r_1$ at rapidity $y = 0$ and
evolve it using the BFKL evolution up to $y=Y$. The number density of dipoles
of size $r_2$ at $Y$ in this dipole, $n(r_1,r_2,Y)$, obeys a completeness
relation
\begin{equation}
n(r_1,r_2,Y) = \int \frac{d^2 r}{2\pi r^2} \ n(r_1,r,Y/2) \ n(r,r_2,Y/2) \ 
\label{eq:nlf}
\end{equation}
where on the right hand side the rapidity evolution is separated in two
successive steps, $y=0 \to y=Y/2 \to y=Y$. With
\begin{equation}
T(r_1,r_2,Y) \simeq c \ \alpha_s^2 \ r_2^2 \ n(r_1,r_2,Y) 
\label{eq:Tn}
\end{equation}
eq.(\ref{eq:nlf}) can be approximately rewritten in terms of the $T$-matrix as
\begin{equation} 
  \left(\frac{1}{r_2^2} \ T(r_1,r_2,Y)\right) \simeq \ \frac{1}{2 c \alpha_s^2}\ \int d\rho \
  \ \left ( \frac{1}{r^2} \ T(r_1,r,Y/2) \right) \ \left( \frac{1}{r_2^2} \ 
    T(r,r_2,Y/2) \right) 
\label{eq:Tlf}
\end{equation}
where $\rho = \ln(r^2_0/r^2)$. In Ref.~\cite{Mueller:2004se} it was realized
that the above completeness relations, or, equivalently, the Lorentz
invariance, is satisfied by the BK evolution only by violating unitarity
limits. This can be illustrated as follows: Suppose that $r_2$ is close to the
saturation line, $r_2 \simeq 1/Q_s(Y/2)$, so that the left hand side of
Eq.(\ref{eq:Tlf}) is large.  On the right hand side of Eq.(\ref{eq:Tlf}) it
turns out that $T(r_1,r,Y/2)/r^2$ is typically very small in the region of
$\rho$ which dominates the integral. This means that $T(r,r_2,Y/2)/r_2^2$ must
be typically very large and must violate unitarity, $T(r,r_2,Y/2) \gg 1$, in
order to satisfy (\ref{eq:Tlf}).

The simple procedure used in Ref.~\cite{Mueller:2004se} to solve the above
problem was to limit the region of the $\rho$-integration in Eq.(\ref{eq:Tlf})
by a boundary $\rho_2(Y/2)$ so that $T(r,r_2,Y/2)/r_2^2$ would never violate
unitarity, or $T(r_1,r,Y/2)/r^2$ would always be larger than $\alpha_s^2$. 
The main consequence of this procedure, i.e., BK evolution plus boundary
correcting it in the weak scattering region, is the following scaling behaviour of
the $T$-matrix near the unitarity limit
\begin{equation}
T(r,Y) = T\left(\frac{\ln(r^2Q_s^2(Y))}{\alpha_s Y/(\Delta\rho)^3}\right)
\label{eq:Trbmf}
\end{equation}
and the following energy dependence of the saturation momentum
\begin{equation}
Q_s^2(Y) = Q_0^2 \ \mbox{Exp}\!\left[\frac{2\alpha_s N_c}{\pi}
\frac{\chi(\lambda_0)}{1-\lambda_0} Y \left(1-\frac{\pi^2
    \chi''(\lambda_0)}{2(\Delta\rho)^2 \chi(\lambda_0)}\right)\right]
\label{eq:Qrbmf}
\end{equation}
with
\begin{equation}
\Delta\rho = \frac{1}{1-\lambda_0} \ln\frac{1}{\alpha_s^2} +
\frac{3}{1-\lambda_0} \ln \ln \frac{1}{\alpha_s^2} + \mbox{const.} \ .
\end{equation}
Eq.(\ref{eq:Trbmf}) shows a breaking of the geometric scaling which was the
hallmark of the BK equation (cf. Eq.(\ref{eq:gs})) and Eq.(\ref{eq:Qrbmf}) shows the correction to
the saturation momentum due to the evolution beyond the mean field
approximation (cf. Eq.(\ref{eq:Qrmf})).

\subsubsection{Statistical physics - high density QCD correspondence}
%
%Another access to small-$x$ physics beyond teh mean field approximation,
%inspired by the dynamics of reaction-diffusion processes in statistical
%physics, was outlined in Ref.~~\cite{Iancu:2004es}.
The high energy evolution can be viewed also in another way which is inspired
by dynamics of reaction-diffusion processes in statistical
physics~\cite{Iancu:2004es}. To show it, let's consider an elementary target
dipole of size $r_1$ which evolves from $y=0$ up to $y=Y$ and is then probed
by an elementary dipole of size $r$, giving the amplitude
$\bar{T}(r_1,r,Y)$. It has become clear that the evolution of the target
dipole is {\em stochastic} leading to random dipole number realizations inside
the target dipole at $Y$, corresponding to different events in an experiment.
The physical amplitude, $\bar{T}(r_1,r,Y)$, is then given by averaging over
all possible dipole number realizations/events, $\bar{T}(r_1,r,Y) = \langle
T(r_1,r,Y)\rangle$, where $T(r_1,r,Y)$ is the amplitude for dipole $r$
scattering off a particular realization of the evolved target dipole at $Y$.

The mean field descripton breaks down at low target dipole occupancy due to
the {\em discreteness and the fluctuations of dipole numbers}. Because of
discreteness the dipole occupancy can not be less than one for any dipole
size.  Taking this fact into account by using the BK equation with a cutoff
when $T$ becomes of order $\alpha^2_s$~\cite{Iancu:2004es}, or the occupancy
of order one (see Eq.(\ref{eq:Tn})), leads exactly to the same correction for
the saturation momentum as given in Eq.(\ref{eq:Qrmf}).  The latter cutoff is
essentially the same as, and gives a natural explanation of, the boundary
used in Ref.\cite{Mueller:2004se} and briefly explained in the previous section.

%The physical amplitude is, of course, given by averaging over all possible
%gluon number realizations/events,
%%
%\begin{equation}
%\bar{T}(r_1,r_2,Y) = \langle T(r_1,r_2,Y)\rangle \ , 
%\end{equation}
%%
%where $T(r_1,r_2,Y)$ denotes the amplitude for dipole $r_2$ scattering
%off a particular realization of the evolved target dipole at $Y$. 

%The mean field decription should be a good approximation so long as the dipole
%occupancy is large compared to one in the evolved target dipole. At low dipole
%occupancy, the {\em discreteness} of the dipole number becomes important,
%especially the fact that dipole occupancy can not be less than one for any
%dipole size. The BK equation with a cutoff when $T$ becomes of size
%$\alpha^2_s$, or the occupancy of order one (see Eq.(\ref{})), leads to
%exactly the same correction for the saturation momentum as given in
%Eq.(\ref{}).  The latter cutoff is essentially the same as the boundary used
%in Ref.\cite{} and briefly explained in the previous section.

The dipole number fluctuations in the low dipole occupancy region result in
fluctuations of the saturation momentum from event to event, with the strength
\begin{equation}
\sigma^2 =  \langle \rho_s^2 \rangle
- \langle \rho_s \rangle^2 = \mbox{const.} \  \frac{\alpha_s
  Y}{(\Delta\rho)^3} \ 
\end{equation}
extracted from numerical simulations of statistical models. The averaging over
all events with random saturation momenta, in order to get the physical
amplitude, causes the breaking of the geometric scaling and replaces it by the
scaling law
\begin{equation}
\langle T(r,Y) \rangle = T\left(\frac{\ln(r^2 Q_s^2(Y))}{\sqrt{\alpha_s
      Y/(\Delta\rho)^3}}\right) \ .
\label{eq:Tqs}
\end{equation}
This equation differs from Eq.(\ref{eq:Trbmf}) since Eq.(\ref{eq:Trbmf})
misses dipole number fluctuations.

\subsubsection{Pomeron loop equations}

It was always clear that the BK equation does not include fluctuations.
However, it took some time to realize that also the Balitsky-JIMWLK equations
do miss them. As soon as this became clear~(first Ref. in \cite{Iancu:2004iy}), the so-called
Pomeron loop equations~\cite{Mueller:2005ut,Iancu:2004iy} have been
constructed, aiming at a description of fluctuations. They can be written
(schematic way, transverse dimensions ignored) as a stochastic equation of
Langevin-type,
\begin{equation}
\partial_Y T  = \alpha_s \left[T  - T T + \alpha_s\ \sqrt{T} \ \nu\right] 
\label{eq:Le}
\end{equation}
or as a hierarchy of coupled equations of averaged amplitudes, the first two
of them reading
\begin{eqnarray}
\partial_Y \langle T \rangle &=& 
\alpha_s \left[ \langle T \rangle  - \langle T T \rangle \right]\nonumber \\
\partial_Y \langle T T \rangle &=&  
\alpha_s \left[ \langle T T \rangle - \langle T\,T\,T \rangle + 
\alpha^2_s \langle T \rangle \right] \ .
\label{eq:hie}
\end{eqnarray}
The last term in Eq.(\ref{eq:Le}), containing a non-Gaussian noise $\nu$, is
new as compared with the BK equation and accounts for the fluctuations in the
dipole number.  Eq.(\ref{eq:hie}) reduces to the BK equation only in the mean field
approximation, i.e., if $\langle T\,T \rangle = \langle
T\rangle \langle T\rangle$. The hierarchy in Eq.(\ref{eq:hie}), as
compared with the Balitsky-JIMWLK hierarchy, involves in addition to linear
BFKL evolution and pomeron mergings, also pomeron splittings, and therefore
{\em pomeron loops}.  The three pieces of evolution are represented by the
three terms in the second equation in Eq.(\ref{eq:hie}), respectivelly, in the
case where two dipoles scatter off a target.

%The complexity of the present evolution equations has hindered  analytic, as
%well as numeric, solutionThe present evolution equations have not been solved so farDue to the complexity of the evolution equations%

It isn't yet clear at which energy fluctuation effects start becoming
important. The results shown in the previous sections, Eq.(\ref{eq:Qrbmf}) and
Eq.(\ref{eq:Tqs}), are valid at asymptotic energies. A solution to the
evolution equations, which is not yet available because of their complexity,
would have helped to better understand the subasymptotics. However, using the
methods outlined in the previous subsections, phenomenological consequences of
fluctuations in the fixed coupling case have been studied, for example for DIS
and diffractive cross sections~\cite{Hatta:2006hs}, forward gluon production in hadron-hadron
collisions~\cite{Iancu:2006uc} and for the nuclear modification factor $R_pA$~\cite{Kozlov:2006qw}, in
case fluctuations become important in the range of LHC energies.

%The new scaling behaviour
%(diffusive scaling) for DIS cross sections, diffractive cross sections and
%multiplicities in hadron-hadron collisions has been studied in~\cite{} and a
%new phenomenon originating from fluctuations, ``total gluon shadowing''for
%$R_pA$ in the fixed coupling, was found in~\cite{}. 

% ****************************************************************************
% BIBLIOGRAPHY AREA
% ****************************************************************************

\begin{footnotesize}
% IF YOU DO NOT USE BIBTEX, USE THE FOLLOWING SAMPLE SCHEME FOR THE REFERENCES
% ----------------------------------------------------------------------------

% ----------------------------------------------------------------------------

% IF YOU USE BIBTEX,
% - DELETE THE TEXT BETWEEN THE TWO ABOVE DASHED LINES
% - UNCOMMENT THE NEXT TWO LINES AND REPLACE 'Name_Of_Your_BibFile'

%\bibliographystyle{unsrt}
%\bibliography{Name_Of_Your_BibFile}
% example of Name_Of_Your_BibFile.bib
% @Article{Turcato:2006ch,
%      author    = "Turcato, M.",
%  collaboration = "ZEUS and H1",
%      title     = "Lepton flavour violation and charmonium physics at HERA",
%      journal   = "Nucl. Phys. Proc. Suppl.",
%      volume    = "162",
%      year      = "2006", 
%      pages     = "283-287",
%      SLACcitation  = "%%CITATION = NUPHZ,162,283;%%"
% }
% 
% @Unpublished{Gogitidze:2007du,
%      author    = "Gogitidze, N.",
%  collaboration = "H1", 
%      title     = "Prompt photons and particle momentum distributions at
%                   HERA", 
%      year      = "2007",
%      note    = "hep-ex/0701033",
%      SLACcitation  = "%%CITATION = HEP-EX 0701033;%%"
% }

\end{footnotesize}

% ****************************************************************************
% END OF BIBLIOGRAPHY AREA
% ****************************************************************************

\end{document}